\documentclass[]{pasj02} 
\usepackage[switch,mathlines]{lineno} 

\jyear{2026}
\Received{}
\Accepted{}

\usepackage{comment}
\usepackage{natbib}
\usepackage{enumitem}
\usepackage{color}

\begin{document} 

\title{Institutional broadening of the Japanese astronomy and astrophysics research community during 1996--2025}

\author{Hideaki \textsc{Fujiwara}\orcid{0000-0001-6536-8656} }
\affil{Frontier Research Institute for Interdisciplinary Sciences (FRIS), Tohoku University, 6-3 Aramaki-aza Aoba, Aoba-ku, Sendai, Miyagi 980-8578, Japan}
\email{hideaki@fris.tohoku.ac.jp}


\maketitle

\begin{abstract}
Using bibliometric data from publication records, we investigate the institutional broadening of Japanese astronomy and astrophysics over 1996--2025. The number of Japan-affiliated publications increased from 588 to 2,156, and the number of unique participating Japanese institutions increased from 123 to 227, indicating a nominal broadening of the institutional base. However, institution--publication links remained concentrated among leading institutions, and the effective number of institutions, a frequency-weighted diversity measure, increased more gradually than the number of unique participating institutions. These results show that Japanese astronomy and astrophysics broadened institutionally, but not through a complete equalization of institutional participation. The observed change is best interpreted as an expansion of the participating institutional base around a persistent core of leading institutions.
\end{abstract}


\section{Introduction}\label{sec:1}

Bibliometric analysis provides quantitative methods for examining research production, impact, and collaboration across fields and over time. When applied with appropriate caution, bibliometric indicators can complement qualitative assessment and help contextualize research activity at the levels of projects, journals, institutions, and countries \citep[e.g.,][]{Moed2010,Bornmann2012,Waltman2016}. In astronomy and astrophysics, publication and citation data have been used to characterize journals, facilities, collaborations, and national or institutional research profiles \citep[e.g.,][]{Abt1981,Abt1996}. For Japanese astronomy, recent bibliometric studies have established baseline views of research activity. These studies illustrate that bibliometric data can provide useful historical and sociological perspectives on Japanese astronomical research \citep{Fujiwara2025,Fujiwara2026}.

One important aspect of such research activity is the institutional base of the community. Astronomy and astrophysics are supported by universities, national research institutes, observatories, space agencies, and other organizations. Changes in the set of institutions appearing in publication records can therefore provide insight into how broadly research activity is distributed across the community. From the perspective of the cultural and institutional history of Japanese astronomy, documenting such changes is also important as a quantitative record of how the research community has evolved.

In this paper, we examine whether the institutional base of Japanese astronomy and astrophysics broadened over 1996--2025. Using bibliometric data from publication records, we identify Japan-affiliated publications in astronomy and astrophysics and the Japanese institutions appearing in these records. We then quantify changes in publication output, participating institutions, institutional concentration, and frequency-weighted institutional diversity. Through these indicators, we ask whether the increase in participating institutions represents a broadening of the institutional base and whether such broadening was accompanied by a reduction in institutional concentration.

\section{Data and methodology}\label{sec:2}
\subsection{Dataset construction}\label{ssec:21}

As briefly reviewed in \citet{Fujiwara2025}, several major bibliographic databases are widely used for literature analysis in astronomy and astrophysics, including the NASA Astrophysics Data System (ADS)\footnote{$\langle$https://ui.adsabs.harvard.edu$\rangle$.}. In this study, we use Elsevier's Scopus\footnote{$\langle$https://www.scopus.com/$\rangle$.} as the primary source because, through its analytics platform SciVal\footnote{$\langle$http://scival.com/$\rangle$.}, it provides standardized subject classifications and allows publication lists to be filtered by Countries/Regions. This functionality is useful for constructing a reproducible list of Japan-affiliated publications within the Astronomy and Astrophysics subject area.

We first searched SciVal for all publications classified within the subject area ``Astronomy and Astrophysics'' (All Science Journal Classifications (ASJC) code 3103) over the year range 1996--2025. This search identified 499,051 publications in the field over the study period. We then applied the ``Countries/Regions’’ filter to this publication set, selecting Japan. This step identified publications that included at least one author affiliated with a Japanese institution. The resulting Japan-affiliated publication set contained 37,783 publications. We downloaded this publication list from Scopus via SciVal together with the available metadata, including publication year, bibliographic information, author and affiliation information, and SciVal Institution IDs. The Scopus database used for retrieval via SciVal was last updated on June 24, 2026. Because data retrieval was conducted several months after the end of 2025, the 2025 publication records were considered sufficiently indexed for the purposes of this longitudinal analysis.

To construct the corresponding list of Japanese institutions, we extracted all SciVal Institution IDs appearing in the Japanese publication list. For each Institution ID, we retrieved institutional metadata using the SciVal Institution Lookup API V1. The returned metadata include the institution name, country, region, sector, and name variants. We then selected institutions whose country was Japan. This procedure yielded 535 Japanese institutions identified by SciVal Institution IDs. The resulting institution list, hereafter referred to as the Japanese institution list, was used as the basis for all institution-level analyses in this study.

Institutional names and identifiers were used as provided by Scopus/SciVal at the time of data retrieval. Thus, when institutions had been merged, reorganized, or renamed during the study period, earlier publications may be associated with institutional entities as normalized by Scopus/SciVal at the time of retrieval. We did not manually reconstruct historical institutional entities or reassign publications to predecessor institutions.

\subsection{Institutional counting and indicators}\label{ssec:22}

The basic unit of institutional counting in this study is the SciVal Institution ID. This identifier represents a SciVal-normalized institutional entity and does not necessarily correspond one-to-one to a legal organization, university corporation, campus, or research institute. A known limitation of this type of institution-level analysis is that parent-level umbrella organizations and subordinate institutes may appear as separate entities. To reduce double counting caused by such hierarchical structures, we excluded three parent-level entities from the Japanese institution list before computing institution-level indicators: Japan Aerospace Exploration Agency (SciVal Institution ID: 208044), National Institutes of Natural Sciences (208110), and Research Organization of Information and Systems (706618). Subordinate entities with their own Institution IDs were retained. We did not manually merge institutional hierarchies. The results should therefore be interpreted as describing SciVal-normalized institutional entities after excluding selected parent-level umbrella institutions, rather than legal institutional units.

For each publication in the Japanese publication list, we identified the Japanese institutions appearing in the publication record by matching the Institution IDs in the publication metadata with the Japanese institution list. When multiple authors in the same publication were associated with the same Institution ID, or when the same Institution ID appeared more than once within a publication record, that Institution ID was counted only once for that publication.

We define an institution--publication link as a unique pair consisting of one publication and one Japanese institution appearing in that publication. If a publication included multiple Japanese institutions, it contributed one institution--publication link to each of those institutions. For each year $t$, we computed the number of publications, $N_{\rm pub}(t)$, the number of unique Japanese institutions, $N_{\rm inst}(t)$, and the number of institution--publication links for institution $i$, $L_i(t)$. The total number of Japanese institution--publication links in year $t$ is
\begin{equation}
L_{\rm tot}(t) = \sum_i L_i(t).
\end{equation}

For descriptive purposes, Japanese institutions were also grouped by the sector metadata obtained from the Institution Lookup API. The sector field was used as returned by the API, without manual reclassification. The sector classification should therefore be interpreted as database metadata rather than as an official Japanese institutional classification. Hereafter, unless otherwise noted, ``institutions'' refers to Japanese institutions identified by Institution IDs after excluding the selected parent-level umbrella institutions.

To quantify the concentration of institutional participation, we calculated the annual share of Japanese institution--publication links accounted for by the top $k$ institutions. For each year, institutions were ranked by $L_i(t)$, and the top-$k$ share was defined as
\begin{equation}
S_k(t) =
\frac{\sum_{i \in {\rm top}\ k} L_i(t)}
     {L_{\rm tot}(t)}.
\end{equation}
We computed this measure for $k = 5$, 10, 15, and 20.

As a complementary measure of institutional diversity, we computed the Shannon entropy \citep{Shannon1948} of the annual institution--publication link distribution and converted it into an effective number of institutions following the Hill-number framework \citep{Hill1973}. For each year, the link share of institution $i$ is
\begin{equation}
p_i(t) =
\frac{L_i(t)}{L_{\rm tot}(t)}.
\end{equation}
The Shannon entropy is then
\begin{equation}
H(t) =
-\sum_i p_i(t)\log_2 p_i(t),
\end{equation}
and the corresponding effective number of institutions is
\begin{equation}
N_{\rm eff}(t) =
2^{H(t)}.
\end{equation}
The effective number can be interpreted as the number of equally participating institutions that would produce the observed entropy. This is equivalent to the order-one Hill number and provides a frequency-weighted measure of institutional diversity, complementing the nominal count of unique institutions.

\section{Results}\label{sec:3}

\subsection{Growth of publication output and participating institutions}\label{ssec:31}

\begin{figure}
 \begin{center}
  \includegraphics[width=90mm]{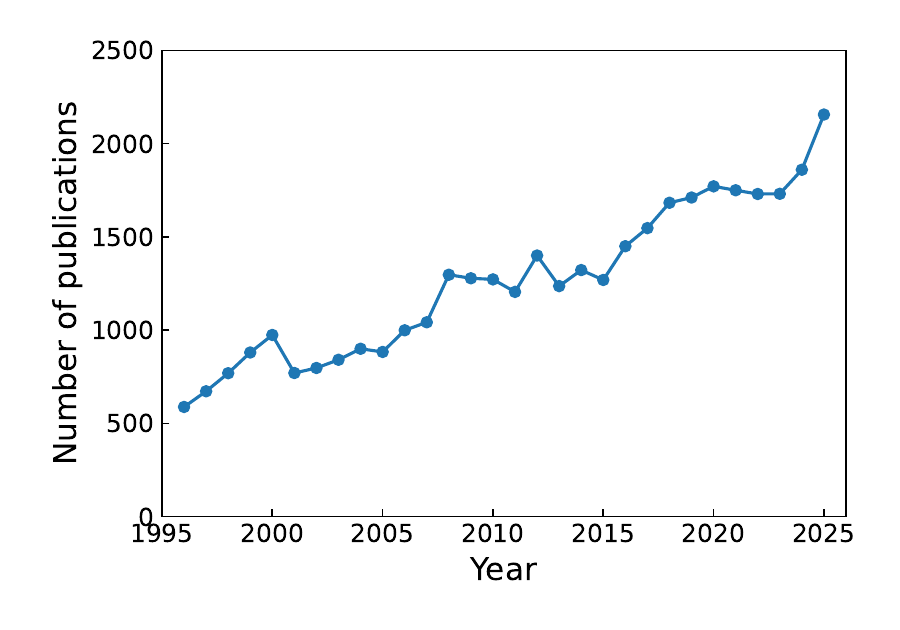}
  \includegraphics[width=90mm]{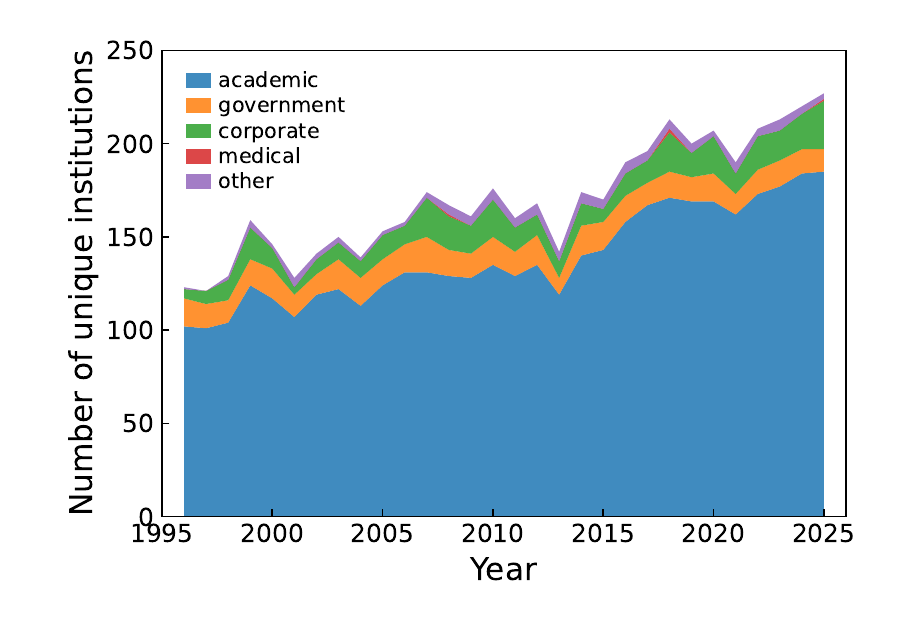}
 \end{center}
   \caption{Basic trends in Japanese participation in astronomy and astrophysics publications, 1996--2025. (Top) Number of publications in the Japanese publication list. (Bottom) Number of unique Japanese institutions appearing in the publications, classified by SciVal sector. Selected parent-level umbrella institutions were excluded from the Japanese institution list before institution-level counting. SciVal Institution IDs are used as the unit of institutional counting; therefore, the counts refer to SciVal-normalized institutional entities rather than legal organizations. {Alt text: Two vertically stacked graphs. The top panel is a line graph with the x-axis showing years from 1996 to 2025 and the y-axis showing the number of publications, ranging from 0 to about 2500. The line increases overall from the late 1990s to the 2020s. The bottom panel is a stacked-area graph with the x-axis showing years from 1996 to 2025 and the y-axis showing the number of unique institutions, ranging from 0 to about 250. The stacked areas represent SciVal sectors, with the academic sector forming the largest component throughout the period.}}
  \label{fig:basic_trends}
\end{figure}

Figure~\ref{fig:basic_trends} summarizes the basic annual trends in Japan-affiliated astronomy and astrophysics publications and the corresponding Japanese institutional participation over 1996--2025. The top panel shows the annual number of publications in the Japanese publication list. The number of publications increased substantially over the study period, from 588 in 1996 to 2,156 in 2025, corresponding to an increase by a factor of 3.7. Although year-to-year fluctuations are present, the long-term trend indicates that Japan-affiliated publication activity in astronomy and astrophysics expanded markedly in volume.

The bottom panel shows the annual number of unique Japanese institutions appearing in these publications, grouped by SciVal sector. The number of participating institutions increased from 123 in 1996 to 227 in 2025, corresponding to an increase by a factor of 1.85. Thus, the increase in publication output was accompanied by a clear increase in the number of Japanese institutions appearing in the publication records, although the growth in participating institutions was more moderate than the growth in publication output.

The sectoral decomposition shows that academic institutions formed the dominant component throughout the period. Academic institutions were already the largest component in the late 1990s, and they also accounted for the largest absolute contribution to the increase in the number of unique participating institutions. Institutions classified in other sectors, including government, corporate, medical, and other categories, also appeared in the publication records, but their contributions to the unique-institution count were smaller than that of the academic sector.

Figure~\ref{fig:basic_trends} shows that Japanese astronomy and astrophysics expanded both in publication output and in the number of participating institutions. The increase in unique institutions indicates a nominal broadening of the participating institutional base. The following subsections examine whether this broadening was accompanied by a reduction in institutional concentration and by an increase in frequency-weighted institutional diversity.

\subsection{Persistent concentration among leading institutions}\label{ssec:32}

\begin{figure}
 \begin{center}
  \includegraphics[width=90mm]{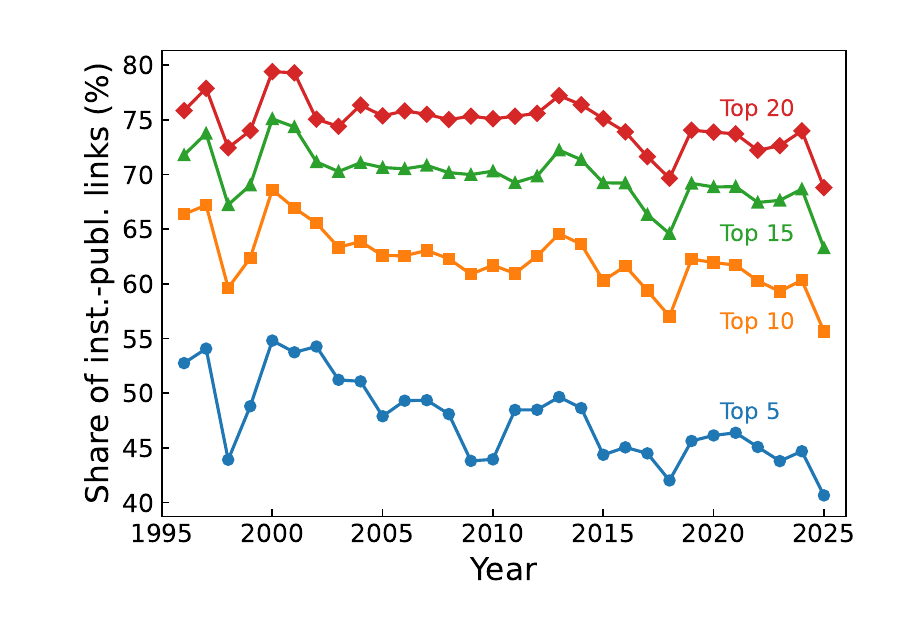}
 \end{center}
\caption{Annual top-$k$ shares of Japanese institution--publication links in astronomy and astrophysics, 1996--2025. For each year, Japanese institutions identified by SciVal Institution IDs were ranked by the number of institution--publication links, and the cumulative shares of the top 5, 10, 15, and 20 institutions were computed. Selected parent-level umbrella institutions were excluded from the Japanese institution list before counting. Persistently high top-$k$ shares indicate that institutional participation remained concentrated among leading institutions despite the broadening of the overall institutional base. {Alt text: A line graph with the x-axis showing years from 1996 to 2025 and the y-axis showing the share of institution--publication links, ranging from about 40\% to about 80\%. Four lines show the annual shares accounted for by the top 5, top 10, top 15, and top 20 Japanese institutions. The top-5 and top-10 lines decline gradually from the late 1990s to the 2020s, while the top-15 and top-20 lines remain higher and change more moderately.}}
  \label{fig:topk_share}
\end{figure}

Figure~\ref{fig:topk_share} shows the annual shares of Japanese institution--publication links accounted for by the top 5, top 10, top 15, and top 20 institutions. In each year, institutions were ranked by the number of institution--publication links, and the cumulative share of the leading institutions was calculated. These curves indicate the degree to which institutional participation was concentrated among institutions with many institution--publication links.

The top-5 and top-10 shares show a gradual decline from the late 1990s to the 2020s. The top-5 share decreased from roughly 50--55\% in the late 1990s and around 2000 to about 45\% in the early 2020s, while the top-10 share declined from about 65\% to about 60\%. This indicates a modest weakening of concentration among the very top institutions.

The declines in the top-15 and top-20 shares are less pronounced. These curves remained at high levels throughout the study period, indicating that a large fraction of institution--publication links continued to be associated with the broader upper tier of leading institutions. Thus, the redistribution of institutional participation appears to have occurred mainly within or around the upper tier, rather than as a complete shift away from leading institutions.

These results show that the institutional base broadened while the distribution of institution--publication links remained uneven. The number of participating institutions increased, and concentration among the very top institutions weakened modestly, but publication participation remained strongly associated with the upper tier of leading institutions. The next subsection examines this pattern using a complementary diversity measure based on the full institution--publication link distribution.

\subsection{Frequency-weighted institutional diversity}\label{ssec:33}

\begin{figure}
 \begin{center}
  \includegraphics[width=90mm]{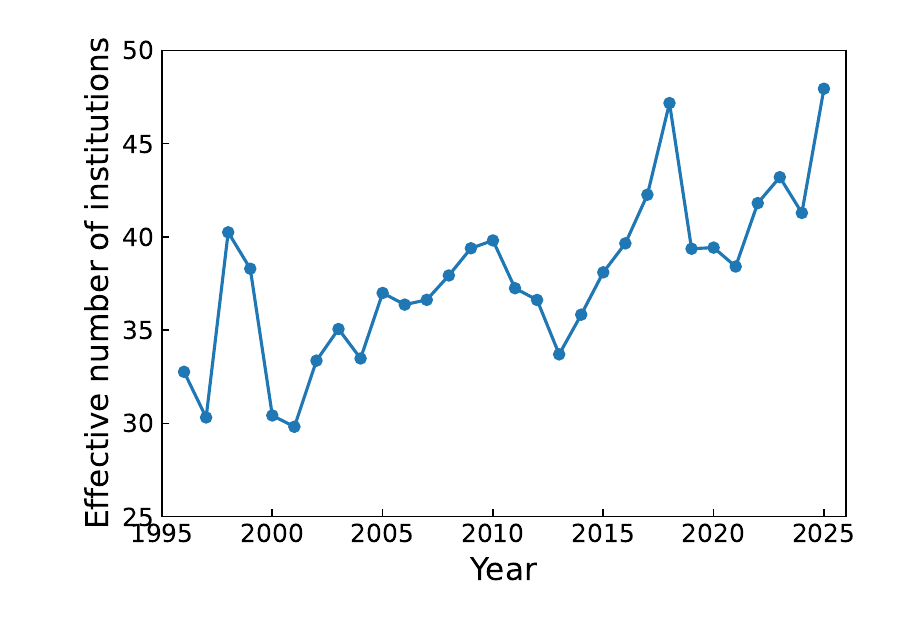}
 \end{center}
\caption{Effective number of Japanese institutions identified by SciVal Institution IDs participating in astronomy and astrophysics publications, 1996--2025. The effective number was derived from the Shannon entropy of the annual institution--publication link distribution. Unlike the nominal count of unique institutions, this measure accounts for the unevenness of institutional participation. Selected parent-level umbrella institutions were excluded from the Japanese institution list before counting. The gradual increase indicates that institutional participation became more diverse in a frequency-weighted sense, although the growth was more modest than the increase in the nominal number of participating institutions. {Alt text: A line graph with the x-axis showing years from 1996 to 2025 and the y-axis showing the effective number of institutions, ranging from about 25 to about 50. The line increases gradually from the late 1990s to the 2020s, with year-to-year fluctuations.}}
  \label{fig:effective_number}
\end{figure}

Figure~\ref{fig:effective_number} shows the effective number of Japanese institutions derived from the Shannon entropy of the annual institution--publication link distribution. This measure complements the nominal count of unique institutions by accounting for the unevenness of institutional participation. If all participating institutions appeared with equal frequency in the publication records, the effective number would approach the nominal number of unique institutions. Conversely, when institution--publication links are concentrated among a smaller number of institutions, the effective number becomes substantially smaller than the nominal count.

The effective number increased over the study period, indicating that institutional participation became more diverse in a frequency-weighted sense. This means that the increase in participating institutions was not purely nominal; the distribution of institution--publication links also became somewhat less concentrated.

However, the increase in the effective number was more modest than the increase in the nominal number of unique institutions shown in the bottom panel of Figure~\ref{fig:basic_trends}. The nominal count of unique institutions increased by a factor of 1.85 between 1996 and 2025, whereas the effective number increased more gradually. This contrast indicates that many of the additional institutions appeared with relatively small numbers of institution--publication links, while a substantial fraction of links continued to be associated with leading institutions.

The effective number therefore provides a compact summary of the main pattern found in this study: the Japanese astronomy and astrophysics research community broadened not only in the nominal sense that more institutions appeared in publication records, but also in the frequency-weighted sense that participation became more institutionally diverse. Nevertheless, because the effective number remained much smaller than the nominal institution count, this broadening should be interpreted as an expansion of participation around a persistent core of leading institutions, rather than as an equalization of institutional participation.

\section{Discussion}\label{sec:4}

The results indicate that the institutional base of Japanese astronomy and astrophysics broadened over 1996--2025, but that this broadening did not correspond to a complete equalization of institutional participation. The number of unique Japanese institutions increased, while institution--publication links remained concentrated among leading institutions. At the same time, the gradual increase in the effective number indicates that the broadening was not merely nominal; institutional participation became more diverse in a frequency-weighted sense. The overall pattern is therefore best described as a broadening of the participating institutional base accompanied by a persistent, though modestly weakened, concentration among leading institutions.

As a contextual check, we compared the growth of unique Japanese institutions with annual counts of Japanese higher education institutions compiled from the School Basic Survey conducted by the Ministry of Education, Culture, Sports, Science and Technology (MEXT) \citep{MEXTSchoolBasicSurvey}. The School Basic Survey data were obtained via the Portal Site of Official Statistics of Japan (e-Stat\footnote{$\langle$https://www.e-stat.go.jp/$\rangle$.}). Table~\ref{tab:growth_indices} summarizes the change between 1996 and 2025, with the index normalized to 1.00 in 1996. The number of unique Japanese institutions increased from 123 to 227, whereas the number of Japanese higher education institutions decreased slightly from 1,236 to 1,162 over the same period. This contrast suggests that the observed broadening was not simply a reflection of an increasing number of higher education institutions in Japan. This comparison should be interpreted as contextual rather than as a strict normalization, because the institution count used in this study and the higher education baseline do not represent the same institutional population. The institution count used in this study includes universities, research institutes, government organizations, companies, and other entities identified through publication records, whereas the higher education baseline is restricted to universities, two-year colleges (Tanki Daigaku), and technical colleges (KOSEN).

\begin{table}
\caption{Growth of publication output, participating Japanese institutions, and Japanese higher education institutions between 1996 and 2025. The index is normalized to 1.00 in 1996. Counts of Japanese higher education institutions were compiled from the School Basic Survey conducted by MEXT and obtained via e-Stat \citep{MEXTSchoolBasicSurvey}. The higher education total is the sum of the five indented categories. Selected parent-level umbrella institutions were excluded from the Japanese institution list before counting.}
\label{tab:growth_indices}
{\centering
\begin{tabular}{lrrr}
\hline
Series & 1996 & 2025 & 2025/1996 \\
\hline
Publications & 588 & 2,156 & 3.67 \\
Unique Institutions & 123 & 227 & 1.85 \\
Higher Education Insts., total & 1,236 & 1,162 & 0.94 \\
\quad National University & 98 & 85 & 0.87 \\
\quad Public University & 53 & 103 & 1.94 \\
\quad Private University & 425 & 624 & 1.47 \\
\quad Two-year College & 598 & 292 & 0.49 \\
\quad Technical College & 62 & 58 & 0.94 \\
\hline
\end{tabular}
\par}
\end{table}

\begin{figure}
 \begin{center}
  \includegraphics[width=90mm]{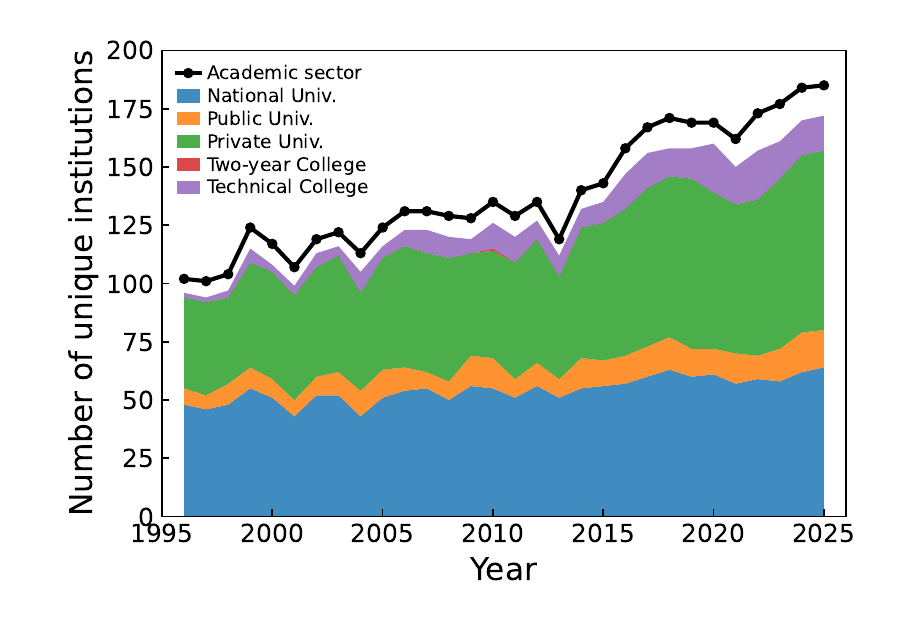}
 \end{center}
\caption{Annual number of unique academic-sector Japanese institutions by institution type, 1996--2025. The black line shows the total number of Japanese institutions classified as academic in the SciVal sector metadata, corresponding to the academic component in the bottom panel of Figure~\ref{fig:basic_trends}. The stacked areas show manually classified higher education categories: national universities, public universities, private universities, two-year colleges, and technical colleges. The comparison indicates that the increase in unique academic-sector institutions was not driven solely by national universities, but also involved public universities, private universities, and technical colleges. The manually classified categories do not exhaust the full SciVal academic sector. {Alt text: A stacked-area plot showing the number of unique academic-sector institutions classified into national universities, public universities, private universities, two-year colleges, and technical colleges. A black line shows the total academic-sector count from SciVal metadata. The plot shows that growth in unique academic-sector institutions includes increases outside national universities.}}
  \label{fig:academic_inst_type}
\end{figure}

The academic sector accounted for the largest component of the increase in unique Japanese institutions in the bottom panel of Figure~\ref{fig:basic_trends}. To examine this component in more detail, we manually classified academic-sector institutions into national universities, public universities, private universities, two-year colleges, and technical colleges. Figure~\ref{fig:academic_inst_type} shows the annual number of unique academic-sector institutions by institution type. The black line shows the total number of Japanese institutions classified as academic in the SciVal sector metadata, corresponding to the academic component in the bottom panel of Figure~\ref{fig:basic_trends}, whereas the stacked areas show the manually classified higher education categories.

Figure~\ref{fig:academic_inst_type} clarifies the composition of the academic-sector broadening. Except for two-year colleges, the number of unique institutions increased in each higher education category. In absolute terms, the increase was especially prominent for private universities, which formed a major component of the growth in unique academic-sector institutions. Technical colleges also showed a considerable relative increase, despite their smaller absolute numbers. By contrast, the increase in national universities was comparatively limited. These patterns indicate that the broadening of the academic sector was not driven primarily by an expansion of national universities, but by the increasing appearance of public universities, private universities, technical colleges, and other non-national-university institutions in the publication records.

A possible explanation for this pattern is the diffusion of astronomy-trained researchers into a wider range of educational institutions, including private universities and technical colleges. Such career movement could have increased the number of institutions appearing in astronomy and astrophysics publication records. However, because the present analysis does not track individual researchers or career histories, this explanation remains speculative and should be examined in future work.

These results indicate that the observed pattern is best interpreted as an expansion of the participating institutional base around a persistent core of leading institutions, rather than as a complete decentralization of institutional participation. In other words, the broadening observed here reflects an increase in the number and variety of institutions appearing in the publication records, while the frequency of participation remained uneven and concentrated among major institutions. The phrase ``institutional broadening'' in this study should therefore be understood as an expansion of the institutional periphery around a still-concentrated core, rather than as a disappearance of the core itself.

Several limitations should be noted. SciVal Institution IDs represent SciVal-normalized institutional entities and do not necessarily correspond one-to-one to legal organizations, university corporations, campuses, or research institutes. Although selected parent-level umbrella institutions were excluded, institutional hierarchies were not manually merged. In addition, SciVal sector metadata and the manually assigned academic institution types should be interpreted as analytical classifications rather than official institutional categories. The manual classification was used only for the contextual decomposition of the academic sector in the Discussion and was not used in the main institution-level indicators. Finally, because the dataset is based on Scopus and SciVal, the results may differ from analyses based on other bibliographic databases such as ADS.

\section{Conclusion}\label{sec:5}

Using Scopus and SciVal data, we examined the institutional broadening of Japanese astronomy and astrophysics over 1996--2025. The number of Japan-affiliated publications and the number of unique participating Japanese institutions both increased, indicating an expansion of the nominal institutional base. In addition, the effective number of institutions derived from the institution--publication link distribution also increased, demonstrating that institutional participation became more diverse in a frequency-weighted sense. Thus, the observed broadening was not merely a numerical increase in institutions appearing in publication records. At the same time, institution--publication links remained concentrated among leading institutions, and the effective number increased more gradually than the nominal institution count. The observed change is therefore best interpreted as a measurable broadening of the participating institutional base around a persistent core of leading institutions, rather than as a complete equalization or decentralization of institutional participation.

\begin{ack}
This work was supported by JSPS KAKENHI Grant Number JP26K03645.
\end{ack}

\end{document}